# Stock mechanics: unification with economy

Çağlar Tuncay
Department of Physics, Middle East Technical University
06531 Ankara, Turkey
caglart@metu.edu.tr

**Abstract**

Associating stock mechanics to real economy, in terms of volume, number of transactions, and cost, i.e. money flow for shares, we obtained the fundamental laws of stock mechanics.

**Unification**

Price ($\chi$) and volume ($\Upsilon$) are basic quantities for shares. For a given number of transactions ($\Lambda$) realized, volume is defined as

$$\Upsilon = \Lambda \chi_{av} \quad , \tag{1}$$

where $\chi_{av}$ is the weighted average price over $\Lambda$. Difference between closures and $\chi_{av}$ becomes negligible in trends averaged over months or years, and we drop the subindex (one may read $\chi$ as $\chi_{av}$, as well). Cost of the process performed over $\Lambda$ many transactions is the money flow, i.e. time (t) derivative of $\Upsilon$,

$$\Phi = d\Upsilon/dt \quad . \tag{2}$$

We have four parameters and two defining equations (Eqs. (1), (2)). So, we have two independent parameters, which may be selected as $\Lambda$, and $\Phi$. We may state that price (and volume) is shaped by $\Lambda$, and $\Phi$.
We may write a differential form of the defining equation for volume (Eq.(1)) as

$$d\Upsilon = \Phi dt = \chi d\Lambda + \Lambda d\chi \quad , \tag{3}$$

which may be called the impact equation of stock mechanics, where some money is utilized to buy (sell) $d\Lambda$ much transactions at a price $\chi$ and rest is utilized to increase (decrease) the price of $\Lambda$ many transactions by $d\chi$. $\Phi dt$ is the cost of the process performed in a period of dt.
The function $\Phi$ fluctuates about zero; what we mean by $\Phi$ in the following sections is the slope of a curve averaging $\Upsilon$ in medium- or long-range. $\Lambda$ is not a well-defined quantity for any index due to inhomogeneity of prices of the shares involved.

*Case 1: Exponential Φ*

With $\Phi=\Phi_0 e^{i\gamma t}$ (i.e., with $\Upsilon=\Upsilon_0 e^{i\gamma t}$, for example) and several forms of $\Lambda$, we obtain a diversity of behavior in price.

*1-a: Constant Λ*

If for a given share (and for an index, roughly) $\Lambda=\Lambda_0$=constant, $\Upsilon$ and $\chi$ increase (or decrease) exponentially with the same exponent, i.e. $\gamma=\beta$, where $\gamma$ and $\beta$ are the exponents of $\Upsilon$ and $\chi$, respectively. The (averaged) price returns and time rate of change of returns is $\gamma\chi$ and $\gamma^2\chi$, respectively, and they vary exponentially. This case may be met, when new buyers continue to come to (or go out of) the market on the condition of survival of positive (or negative) expectation. Thus, old buyers continuously have the good (or bad) chance of selling their shares with some profit (or loss). Old buyers may become new buyers at new prices (of the shares they have sold, or in new ones), and the total number of exchanged shares remains almost constant.

*1-b: Exponential Λ*

For $\Lambda=\Lambda_0 e^{\lambda t}$, we have $\chi=\chi e^{\beta t}$, where $\beta=\gamma-\lambda$. This is an equivalent case to 1-a, provided $\beta$ is shifted from $\gamma$ to $\gamma-\lambda$. Therefore, for exponentially growing volume (and money), whether $\Lambda$ is constant or exponential, price becomes exponential.

In case 1-b, if $\lambda<0$, price increases more rapidly than $\Upsilon$, with $0<\gamma$; and offer may decrease exponentially for exponentially increasing demand. The present stituation may be met if highly positive expectations exist, so that demand highly exceedes offer; there exist more buyers than the booked number of $\Lambda$. Considering the other relevant combinations of values and signs of $\gamma$, and $\lambda$, we may generalize that; $\gamma$ may be called as (daily) demand parameter and $\lambda$ as (daily) offer parameter. The difference of them determines the exponent of price, that of returns, etc.

In the present case, whenever offer parameter exceeds demand one, returns changes sign. As a result, such short-term variations in the sign of the difference of demand and offer parameter may cause the price to fluctuate in short-range, without altering longer-range trends.

*1-c: Linear Λ*
There may be some terms, when $\Lambda$ varies linearly, or if $\lambda<<1$ in absolute value, then $\Lambda=\Lambda_0 e^{\lambda t}$ may be aprroximated by $\Lambda=\Lambda_0(1+\lambda t)$. And we have, $\chi=\chi_0 e^{\gamma t}/(1+\lambda t)$, with $\chi_0=\Upsilon_0/\Lambda_0$. In this stituation we have very different solutions for price depending on the sign of $\lambda$ together with the sign and value of the ratio $\gamma/\lambda$. If $0<\lambda$, denominator in $\chi$ never gets zero for $0<t$, then we have nearly exponential growth of $\chi$, provided $0<\lambda/\gamma \cong 1$. We have steeper growth in price, for $0<\lambda/\gamma<<1$, and for $1<<\lambda/\gamma$, we have nearly exponential decay of $\chi$, which may be attributed to over excess of demand with respect to supply. If $\lambda<0$, we have the denominator zero at $t=-1/\lambda$, and we have singularity in price. Depending on signs and values of $\gamma$ and these of the ratio $\lambda/\gamma$, price displays a wide diversity of excursions.

*Case 2: Sinusoidal Φ*

If voluminous buyings and sellings are being performed alternatively, especially when there is absence of strong (positive or negative) expectation and future is not clearly foreseen; public interest towards the market may cease. And also, market makers may not be wishing to invest their money within the market, they may be prefering short- and medium-range trading, instead.

*2-a: Constant Λ*

Utilization of $\Upsilon=\Upsilon_0 e^{i\gamma t}$ to describe the volume, and substituation of it into Eq. (3) after differentiation, yields

$$i\gamma\Upsilon_0 e^{i\gamma t} = \Lambda_0 d\chi/dt \quad , \tag{5}$$

where $\Lambda_0$ is the constant value for Λ. Price comes out as oscillating sinusoidally with the frequency of volume. In other words, demand parameter becomes as equal to the scalar factor of the purely imaginary exponent of price.

*2-b: Sinusoidal Λ*

If Λ is also sinusoidal; we may utilize $\Lambda=abs(\Lambda_0 e^{i\lambda t})$, where we have absolute value (abs) since negative number of transaction does not exist. Price comes out again sinusoidal, but its exponent changes into $i(\gamma-\lambda)$, and the frequency of the oscillation changes accordingly.

Therefore, in cases 2-a and 2-b, returns and rate of returns are also sinusoidal with the same angular frequency of β, but they are out of phase with respect to the price by a quarter- and a half-period, respectively. In other words, returns are maximum when the price is far from its extrema, and vice versa.

*Case 3: Linear Φ, Constant Λ*

Exponential Φ may be treated as linear for short terms, if the exponent is small. Besides, there may exist some time terms, where the variation in Φ may be approximated linearly, i.e. when the volume varies almost quadratically. For such cases, $\Phi=\Phi_0+\phi t$ may be utilized in Eq. (3). Now, Eq. (3) simplifies a lot.

$$\Phi_0+\phi t=\Lambda_0 d\chi/dt \quad , \tag{6}$$

where $\Lambda_0$ is the constant value for Λ. Returns ($d\chi/dt$) change with time, and they increase (decrease) if $0<\phi$ ($\phi<0$). As a result, the trajectory of price, as a solution of Eq.(6) will be quadratic in time, i.e., parabolic. We will have concave parabola if $0<\phi$, and have convex parabola $\phi<0$, since $\Lambda_0$ is positive by definition, and accordingly returns (as gains and losses) will decrease towards the extremum, and will increase afterwards.

*Case 4: Constant Φ*

Due to survival of positive (negative) expectation in the market, flow of money towards (away from) the market may be constant, $\Phi=\Phi_0$. The money may be spent to buy the booked transactions at a constant price (by squeezing it), it may be spent to increase linearly the price of a constant number of Λ, etc. In any of the relevant stituations we have

$$\Upsilon_0 + \Phi_0 t=\chi\Lambda \quad . \tag{7}$$

If furthermore $\Phi_0=0$, (in absence of expectation or due to changing conditions in market) tendency of the investors (traders) may not be alligned, and herding may be absent. Furthermore, the market makers and players may not wish yet to keep money in the market, (as also described in Case-2.) then we have the price inversely proportional to Λ.

Depending on whether $\Lambda$ is exponential, sinusoidal, linear, inversely linear etc., we have a diversity of price excursion. Of course, there are very many other $\Phi$-$\Lambda$ cases else than the ones pronounced above, taking place within market, which are kept out of the context of the present brief report, for the sake of simplicity.

We obtain a force equation to cover all the possible cases in market, by taking the derivative of money, i.e., the second derivative of volume

$$d\Phi/dt = \Lambda a + 2v d\Lambda/dt + \chi d^2\Lambda/dt^2 \quad , \tag{8}$$

which is nothing but the 2$^{nd}$ of Newton's Laws of classical mechanics with time dependent mass $\Lambda$, and whenever $\Lambda$ may be considered as constant, it may be taken as unity, on the behalf of changing units of the relevant quantities. In Eq. (8), v stands for the speed, (v=d$\chi$/dt) and, a stands for the the acceleration (a=d$^2\chi$/dt$^2$) of price.

**Conclusion**

We combined $\Phi$ and $\Lambda$ with price in impact (Eq. (3)), and in force (Eq. (8)) equations, which are the same as these equations of classical mechanics for particle with variable mass. For terms with constant number of transactions processed ($\Lambda$), we have the force as proportional to the acceleration a, (we have the impact as proportional to speed v, linear momentum) of price. In economy terms, speed is price returns, and acceleration is time rate of change of price returns. Potential energy for price is a well-defined quantity, which is obtained in terms of integrating force in usual manner, under the assumption of conservation of total energy. In some cases covered above, (with constant $\Lambda$, which may be taken as unity[1,2]) in exponential (cases: 1-a, -b), sinusoidal (2-a), and parabolic (3) behavior, potential energy becomes as negative definite quadratic, positive definite quadratic, and linear (gravitational) in price, respectively.[2] Moreover, many mechanical aspects, e.g., extrema values of price and durations of the pronounced cases can be estimated in terms of $\Phi$ and $\Lambda$. For example, in gravitational case, duration of the epoch (time of flight) may be given as $t_{flight} = 2\Phi_0/\phi$, where $\Phi_0$ and $\phi$ is defined in the related case, 3. As final remarks these may be phrased that, radical changes in $\Phi$ and $\Lambda$ may change the present character of price and the form of potential energy and the value of total energy. Furthermore, $\Lambda$ (d$\Lambda$/dt) involves a power law.[3]

**Acknowledgement**

**References**

[1] Ç. Tuncay, "Stoch mechanics: predicting recession in S&P500, DJIA, and NASDAQ", Central European Journal of Physics **4** (1), 58–72 (2006). *Preprint*: arXiv:physics/0506098.
[2] Ç. Tuncay, "Stock mechanics: a general theory and method of energy conservation with applications on DJIA", scheduled for Int. J. Mod. Phys. C 17/ issue 6. *Preprint*: arXiv:physics/0512127.
[3] Ç. Tuncay, "Unified economechanics: applications", on process. Many cases, which are kept out of the context of the present brief report, are involved there.